\documentclass[a4paper]{jpconf}
\usepackage{graphicx,amssymb,amsmath,times}

\usepackage{verbatim}
\renewcommand{\etal}{et al}
\newcommand{\ie}{\emph{i.e.}}
\newcommand{\eg}{\emph{e.g.}}

\newcommand{\Hyd}{\textsf{H}}
\newcommand{\He}{\textsf{He}}

\newcommand{\BC}{\textsf{B/C}}

\newcommand{\AMS}{\textsf{AMS}}

\newcommand{\ApJ}{ApJ}
\newcommand{\AeA}{A\&A}
\newcommand{\ApP}{Astropart. Phys}

\renewcommand{\PRL}{PRL}
\newcommand{\MNRAS}{MNRAS}

\newcommand{\JCAP}{JCAP}

\newcommand{\We}{I}
\newcommand{\we}{I}

\def\Journal#1#2#3#4{{#4} {#1} {#2} #3} 

\begin{document}

\title{Diffusive Origin of the Cosmic-Ray Spectral Hardening}
\author{Nicola Tomassetti}
\address{INFN -- Sezione di Perugia, 06122 Perugia, Italy}
\ead{nicola.tomassetti@pg.infn.it}

\begin{abstract}
Recent data from ATIC, CREAM and PAMELA revealed that the energy spectra of cosmic ray (CR) nuclei above 100 GeV/nucleon experience a remarkable hardening with increasing energy. This effect cannot be recovered by the conventional descriptions of CR acceleration and diffusive propagation processes. Using analytical calculations, I show that the hardening effect can be consequence of a spatial change of the CR diffusion properties in different regions of the Galaxy. I discuss the implications of this scenario for the main CR observables and its connections with the open issues of the CR physics.
\end{abstract}

\section{Introduction}

Understanding the origin of the cosmic ray (CR) energy spectrum is
central to astrophysics and has long been the focus of intensive study.
The spectrum of CR nuclei at $\sim$\,10$^{1}$--10$^{6}$\,GeV/nucleon is thought to be the result of 
diffusive shock acceleration (DSA) mechanisms in supernova remnants (SNR), 
followed by diffusive propagation in the interstellar medium (ISM)  \cite{Strong2007}.
The conventional descriptions predict source spectra such as $E^{-\nu}$
for primary nuclei (\eg, \Hyd, \He)  which are steepened as $E^{-\nu-\delta}$ by diffusion.
The data constrain $\nu+\delta\approx$\,2.7 (depending on the element) and $\delta \sim$\,0.2--0.7,
whereby $\nu\sim$\,2.0--2.5. 
In DSA calculations, the source spectral slope should be
$\nu\lesssim$2.2, which implies $\delta \gtrsim$\,0.5.
On the other hand, large values for $\delta$ are at odds with the
anisotropy observations at TeV energies, which favor $\delta \lesssim$\,0.3.
On top of that, recent experiments ATIC-2, CREAM, and PAMELA, reported
a remarkable spectral hardening at energies above $\sim$\,100\,GeV/nucleon which
cannot be explained by conventional mechanisms \cite{Panov2009,Ahn2010,Adriani2011}.
Proposed explanations of this effect deal with
acceleration mechanisms \cite{Biermann2010,Ptuskin2011}, 
nearby SNRs \cite{Thoudam2012}
or multi-source populations \cite{Zatsepin2006,Yuan2011}.
\We{} propose here that the hardening originates from a spatial
change of the CR diffusion properties in the different regions of the
Galactic halo.
The key hypotesis is that the diffusion coefficient $K$ is not
separable into energy and space terms as usually assumed.
As I will show, this hypotesis leads to a pronounced change
in slope for the energy spectra of CR nuclei at Earth, giving a good
description of the present data.
Remarkably, the model proposed here has also positive impacts on several open problems 
in the CR acceleration/propagation physics.

\section{The two-halo model}

\We{} use a one-dimensional inhomogeneus diffusion model for CR transport and
interactions \cite{NT2012}. 
The Galaxy is modeled to be a disc (half-thickness $h\cong$\,100\,pc) 
containing the gas (number density $n\cong$\,1\,cm$^{-3}$) and the CR sources. 
The disc is surrounded by a diffusive halo (half-thickness
$L\cong$\,5\,kpc) with zero matter density. For each stable CR nucleus, the transport equation is:
\begin{equation}\label{Eq::Transport1D}
\frac{\partial N}{\partial t} = \frac{ \partial}{\partial z} \left( K(z) \frac{\partial N}{\partial z} \right) 
-2h \delta(z) \Gamma^{\rm inel} N + 2h\delta(z) Q \,,
\end{equation}
where $N(z)$ is its number density as function of the $z$-coordinate, $K(z)$ is the position-dependent diffusion 
coefficient and $\Gamma^{\rm inel} = \beta c n \sigma^{\rm inel}$ is the destruction rate in the ISM 
at velocity $\beta c$ and cross section $\sigma^{\rm inel}$. 
The source term $Q$ is split into a primary term $Q_{\rm pri}$, from SNRs, 
and a secondary production term $Q_{\rm sec}= \sum_{\rm j} \Gamma_{j}^{\rm spall} N_{\rm j}$,
from spallation of heavier ($j$) nuclei with rate $\Gamma_{j}^{\rm spall}$.
The quantities $N$, $K$, $Q$ and $\Gamma^{\rm inel}$ depend on energy
too. Equation\,\ref{Eq::Transport1D} can be solved in steady-state conditions
($\partial N/\partial t =0$) with the boundary conditions $N(\pm L)=0$ \cite{NT2012}. 
For each CR nucleus, the differential energy spectrum is given by:
\begin{equation}\label{Eq::SolutionVSz}
J(z,E) \equiv \frac{\beta c}{4 \pi}N(z,E) = \frac{\beta c}{4\pi} \frac{Q(E)}{\frac{K_{0}(E)}{h\Lambda(E)} + \Gamma^{\rm
      inel}(E)} \left[ 1 - \frac{\lambda(z,E)}{\Lambda(E)} \right] \,,
\end{equation}
where $K_{0}=K(z$$\equiv$$0)$, $\lambda = K_{0} \int_{0}^{|z|} \frac{dz}{K(z)}$, and $\Lambda=\lambda(z$$\equiv$$L)$.
In CR propagation studies, the diffusion coefficient is usually assumed to be separable as $K(z,E)\equiv f(z)K_{0}(E)$
(in homogeneus models $f(z)\equiv$\,1, whereby $\Lambda=L$).
In these models, the functions $\lambda$ and $\Lambda$ are independent
on energy, so that the predictions at Earth
($z$$=$$0$) are spectrally uninfluenced by the choice of $f(z)$.
Conversely, I employ a non-separable $K(z,E)$, which reflects the existence of
different CR transport properties in the propagation volume, depending on
the nature and scale distribution of the magnetic-field irregularities. 
In fact, while SNR explosions may generate large irregularities in the region near the Galactic plane,
the situation in the outer halo is different because that medium is undisturbed by SNRs.
From these considerations, the authors of Ref. \cite{Erlykin2002}
found that the turbulence spectrum in the 
halo should be flatter than that in the Galactic plane. 
This implies a strong latitudinal dependence for the parameter $\delta$,
which suggests spatial variations of the CR energy spectra. 
Noticeably, new data reported by \textit{Fermi}/LAT on the diffuse $\gamma$-ray
emission at $\sim$\,10--100\,GeV of energy seem to support these suggestions:
the $\gamma$-ray spectra observed near the Galactic plane (latitude
$|b|<$8$^{\circ}$) are found to be harder than those at higher latitudes \cite{Ackermann2012}.
Following the above arguments, \we{} adopt a simple \textit{two-halo model} consisting in two diffusive zones.
The \textit{inner halo} is taken to surround the disk for a typical size $\xi L$ of a few hundred pc ($\xi\sim$\,0.1).
Its medium properties are influenced by SNRs, which produce a steep turbulence spectrum in terms of energy density per wave number, 
$w(\kappa)d\kappa\sim \kappa^{-2+\delta}d\kappa$, presumably close to the Kolmogorov regime $\delta\sim$\,1/3. 
The \textit{outer halo} represents a wider region, $\xi L <|z|<L$, which is undisturbed by SNRs. 
Its turbulence spectrum is driven by CRs and should be flatter. 
For instance, CRs with rigidity spectrum $R^{-\nu}$ (where $R=p/Z$) 
excite turbulent modes of wave-numbers $\kappa\propto 1/R$,
giving a spectrum $\sim$$\kappa^{-2+\delta^{\prime}}$, where 
$\delta^{\prime}\sim \nu-1$ is of the order of the unity.
This situation can be realized by a rigidity dependent diffusion coefficient of the type
\begin{equation} \label{Eq::TwoHaloDiffCoeff}
K(z,R) = 
\begin{cases}
 k_{0}\beta\left(R/R_{0}\right)^{\delta} & \,{\rm\it for\,} |z|< \xi L \,\,({\rm\it inner\, halo}) \, \\
 k_{0}\beta\left(R/R_{0}\right)^{\delta+\Delta} & \,{\rm\it for\,} |z|>\xi L \,\,({\rm\it outer\, halo})  \,,
\end{cases}
\end{equation}
where $k_{0}\cong$\,0.04\,kpc/Myr specifies its normalization at the reference rigidity $R_{0}\equiv$\,5\,GV.  
For $R>R_{0}$, the diffusion coefficient of Eq.\,\ref{Eq::TwoHaloDiffCoeff} 
produces a higher CR confinement in the inner halo (with
$\delta\sim$\,1/3), whereas the outer halo 
(with $\Delta=\delta^{\prime}-\delta\sim$\,0.5 -- 1) 
represents a \textit{reservoir} from which CRs leak out rapidly and can re-enter the inner halo.
The non-separability of $K$ has a remarkable consequence on the model predictions at $z=0$.
It can be understood if one neglects the term $\Gamma^{\rm inel}$ and
takes a source term $Q_{\rm pri} \propto R^{-\nu}$. From
Eq.\,\ref{Eq::SolutionVSz} and using Eq.\,\ref{Eq::TwoHaloDiffCoeff}, one finds
\begin{equation}\label{Eq::SpectrumTwoComponents}
J_{0} \equiv J(z=0) \sim \frac{L }{k_{0}} \left\{ \xi \left( R/R_{0} \right)^{-\nu -\delta} +
  (1-\xi)\left( R/R_{0} \right)^{-\nu -\delta -\Delta} \right\} \,,
\end{equation}
which describes the CR spectrum as a result of two components. 
Its differential log-slope as a function of rigidity reads
\begin{equation}\label{Eq::LogSlope}
\gamma(R) = - \frac{\partial\log J_{0}}{\partial\log R} \approx \nu + \delta +
\frac{\Delta}{1 + \frac{\xi}{1-\xi}\left(R/R_{0}\right)^{\Delta}}  \,,
\end{equation}
which shows a clear transition between two regimes.
In practice the low-energy regime ($\gamma\approx\nu+\delta+\Delta$) is never reached
due to spallation, that becomes relevant and even dominant over escape 
($\Gamma^{\rm inel} \gtrsim \frac{K_{0}}{h\Lambda}$). In this case the log-slope  
is better approximated by $\gamma\sim \nu + \frac{1}{2}(\delta +\Delta)$ \cite{Blasi2012a}.
The hard high-energy regime ($\gamma \approx \nu+\delta$) is determined by the diffusion properties of the inner halo only.
In this limit one has $\Lambda \approx \xi L$. The effect vanishes at all rigidities when passing to the homogeneus limit of 
$\xi\rightarrow 1$ (one-halo) or $\Delta\rightarrow 0$ (identical halos), where one recovers the usual relation $\gamma = \nu + \delta$.
From Eq.\,\ref{Eq::SolutionVSz}, it can be seen that the intensity of the harder component 
diminishes gradually with increasing $|z|$, \ie, the CR spectra at
high energies are steeper in the outer halo, as suggested by the new \textit{Fermi}/LAT data.


\begin{figure}
\begin{center}
\includegraphics[width=0.45\textwidth]{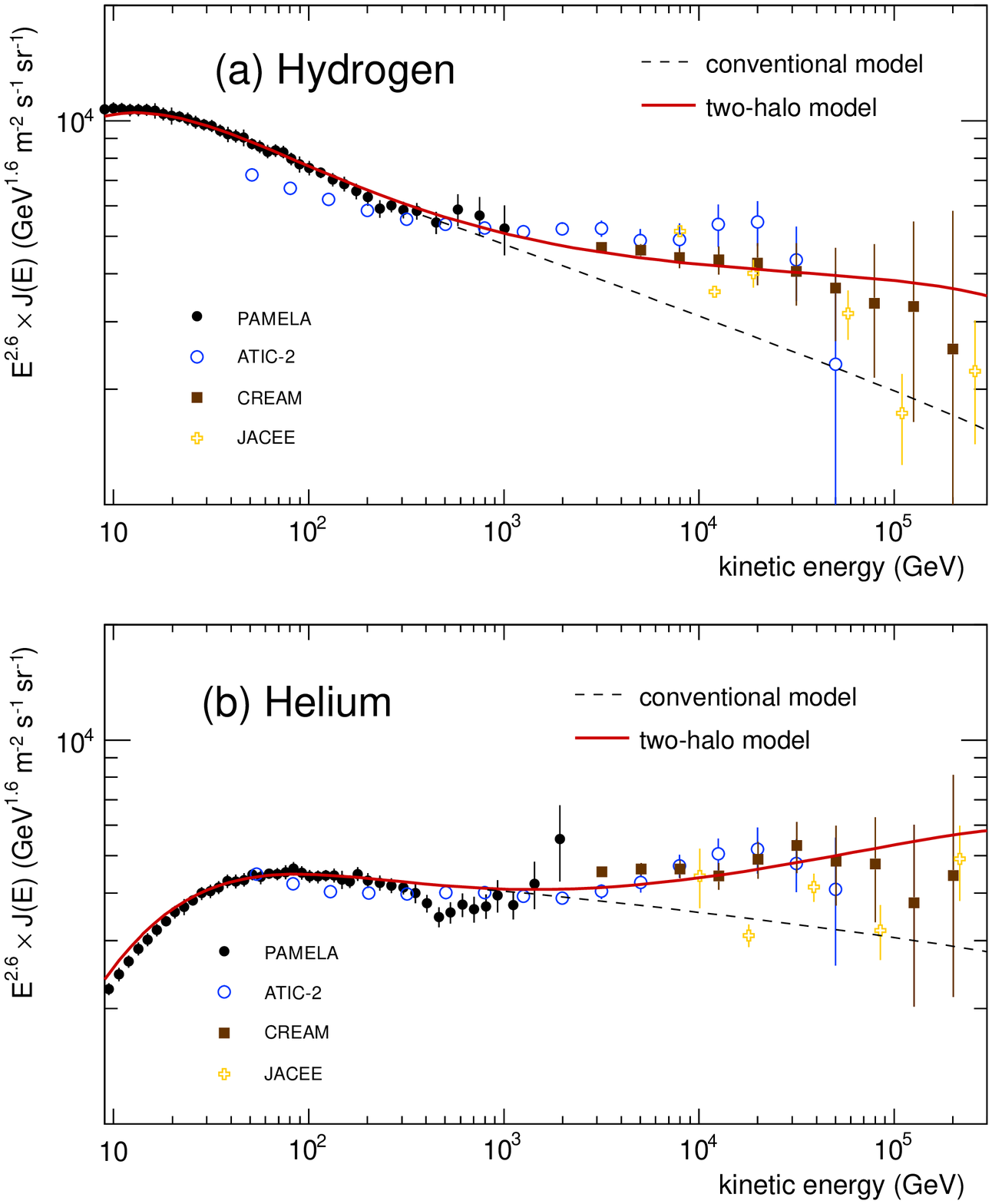}\hspace{2pc}%
\includegraphics[width=0.45\textwidth]{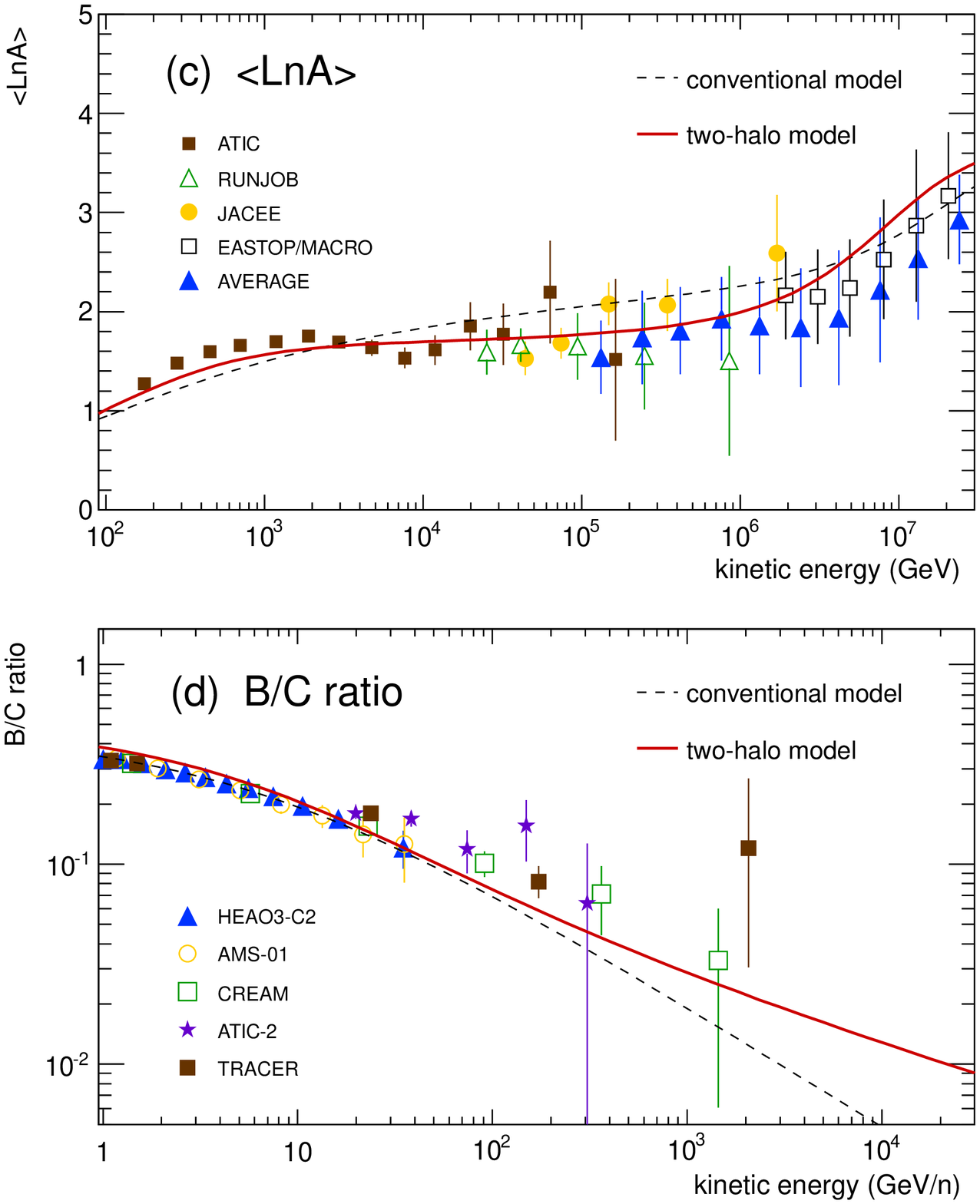}\hspace{2pc}%
\caption{ 
  Left: CR spectra of \Hyd{} and \He. Right: mean logarithmic mass and
  \BC{} ratio. Model calculations are compared with the data
  \cite{Panov2009,Ahn2010,Adriani2011,Asakimori1998, 
  Derbina2005,Aglietta2003,Horandel2004,
  Obermeier2011,Aguilar2010,Panov2007,Engelmann1990,Ahn2008} \label{Fig::ccResults}.
}
\end{center}
\end{figure}

Figure\,\ref{Fig::ccResults} shows the \Hyd{} and \He{} spectra at Earth, from Eq.\,\ref{Eq::SolutionVSz},
at kinetic energies above 10\,GeV.
The implementation of the model follows Refs.\,\cite{NT2012,NTFD2012}.
The source spectra are taken as $Q\propto R^{-\nu}$, with $\nu\cong$\,2.3 for $H$ and $\nu\cong$\,2.16
for \He{} \cite{Malkov2012}.
The two halos are defined by $L\cong$\,5\,kpc and $\xi L\cong$\,0.5\,kpc. 
A Kolmogorov-type diffusion is adopted for the inner halo
($\delta\cong$\,1/3), while $\Delta$ is taken as 0.55,
consistently with Ref.\,\cite{Erlykin2002}. 
Results from the two-halo model (solid lines) are compared with those
from the homogeneus model (dashed lines), which uses $\delta\cong$\,0.6, and with recent CR data.
Below $\sim$\,10\,GeV/nucleon, the solar modulation is apparent and it is described using a \textit{force-field} 
modulation potential $\phi\cong$\,400\,MV \cite{Gleeson1968}. Note that my model may be inadequate 
in this energy region due to approximations. 
Remarkably, my model reproduce well the observed changes in slope at
$\gtrsim$\,100\,GeV, in good agreement with the trend indicated by the data.
It should be noted, however, that the sharp spectral structures suggested by the PAMELA
data at $\sim$\,300\,GeV cannot be recovered. 
Figure\,\ref{Fig::ccResults}c shows the mean logarithmic mass,
$\langle ln(A)\rangle$, which is described well by the two-halo model.
Figure\,\ref{Fig::ccResults}d shows the \BC{} ratio from
the two-halo model, which is predicted to harden as $\propto \Lambda(E)/K_{0}(E)$,
while its low-energy behavior is similar to that of the
homogeneus model (with $\delta=$\,0.6 in the whole halo).
Interestingly, a slight flattening for the \BC{} ratio is also
suggested by recent data, \eg, from TRACER. 
A multi-channel study of the \AMS{} data will allow to resolutely test these features.
The \BC{} ratio hardening is also connected with the global large-scale
anisotropy amplitude, which should increases as $\eta \propto K_{0}/\Lambda$ \cite{Shibata2004,NT2012}. 
Thus, the two-halo model predicts a gradual
flattening of $\eta$, as indicated by the present data. 
It is also interesting noticing that the anisotropy may be furtherly reduced at all energies 
if one accounts for a proper radial dependence for $K$ \cite{Evoli2012}.
In summary, the model I proposed is potentially able to reconcile a weak energy dependent of the
anisotropy amplitude (as suggested by observations) with relatively hard source
spectra $\nu\approx$\,2.2 (as preferred by the DSA theory), giving
good fits to the \BC{} ratio. 
\We{} recall that plain diffusion models employ $\nu\approx$\,2.2
and $\delta\approx$\,0.5 in the whole halo (which gives a too strong energy dependence for $\eta$),
whereas diffusive-reacceleration models use $\nu\approx$\,2.4 and
$\delta\approx$\,1/3 (which are too challenging for the DSA and require strong
reacceleration to match the \BC{} data).
%

\section{Conclusions}     
\label{Sec::Conclusions}  

I have shown that the spectral hardening observed in CRs may be 
consequence of  a Galactic diffusion coefficient that is not separable
in energy and space terms.
From the model presented here, the hardening arises naturally as a local effect and vanishes gradually in the
outer halo, where the CR spectra are also predicted to be steeper.
This effect must be experienced by all CR nuclei as well as by secondary-to-primary ratios. 
Recent data from \textit{Fermi}/LAT and TRACER seem to support this scenario,
but the change in slope predicted by this model is more gradual than that suggested by PAMELA data.
All these features can be tested with the data forthcoming by \AMS{},
so that the present model will be discriminated against other interpretations. 
This scenario has an interesting impact on the CR physics. 
As shown, a Kolmogorov diffusion for the inner halo
($\delta\sim$\,1/3) is consistent with relatively hard source spectra
($\nu\sim$\,2.2), giving good descriptions of both the \BC{} ratio and the high-energy trend
of the anisotropy amplitude.
\\\\
I acknowledge the support of the \textit{Italian Space Agency} under contract ASI-INFN I/075/09/0.
\\

\end{document}